\documentclass[runningheads]{llncs}

\usepackage{graphicx}
\usepackage{hyperref}
\usepackage{tabularx}

\begin{document}

\title{Information Ecosystem Reengineering via \\ Public Sector Knowledge Representation}
\titlerunning{Information Ecosystem Reengineering}

\author{Mayukh Bagchi}

\authorrunning{Mayukh Bagchi}

\institute{Department of Information Engineering and Computer Science (DISI), \\ University of Trento, Trento, Italy. \\ Institute for Globally Distributed Open Research and Education (IGDORE), Kolkata, India. \\
\email{mayukh.bagchi@igdore.org}}

\maketitle              
\begin{abstract}
Information Ecosystem Reengineering (IER) - the technological reconditioning of information sources, services, and systems within a complex information ecosystem - is a foundational challenge in the digital transformation of public sector services and smart governance platforms. From a semantic knowledge management perspective, IER becomes especially entangled due to the potentially infinite number of possibilities in its \emph{conceptualization}, namely, as a result of \emph{manifoldness} in the multi-level mix of perception, language and conceptual interlinkage \emph{implicit} in \emph{all} agents involved in such an effort. This paper proposes a novel approach - Representation Disentanglement - to disentangle these multiple layers of knowledge representation complexity hindering effective reengineering decision making. The approach is based on the theoretically grounded and implementationally robust ontology-driven conceptual modeling paradigm which has been widely adopted in systems analysis and (re)engineering. We argue that such a framework is essential to achieve explainability, traceability and semantic transparency in public sector knowledge representation and to support auditable decision workflows in governance ecosystems increasingly driven by Artificial Intelligence (AI) and data-centric architectures.


\end{abstract}


\section{Introduction}
\label{S01}
The triad of information sources, services and systems constitute the substratum on which any \emph{information ecosystem} functions \cite{2013-ISSS,2021-KOE}. For example, let us consider the motivating scenario of a public sector information governance institution, e.g., a university knowledge resource centre and within it, the case of its \emph{information services chatbot} (see \cite{2020-DJLIT}). Clearly, the demand for services of such artificial conversational agents are at peak due to several disruptive issues such as, for instance, remote work as a consequence of the recent global health emergency of COVID-19 (see \cite{2020-NPJ}). Such disruptions also, \emph{a fortiori}, entail \emph{information ecosystem reengineering} - the need to reengineer the established information \emph{system} of such agents, with an objective to adapt to newer forms of information \emph{services} factoring in evolving heterogeneous information \emph{sources} \cite{1993-BPR}. While the chatbot example originates in a higher education context, similar information ecosystem reengineering challenges are common across public sector digital services such as e-government portals, citizen support e-services, smart municipal information hubs and participatory civic platforms. These public sector systems form the core of modern digital governance infrastructures, where, effective reengineering can significantly impact service responsiveness, policy alignment and trust in government.

Amongst the numerous existing challenges in information ecosystem reengineering, the present work is a first attempt to alleviate the critical lack of \emph{conceptual modelling support} \cite{2020-GGMCM} in streamlining such decisions (elicited from the detailed study in \cite{1998-REENGG}). The lack of exploitation of conceptual modelling frameworks in information ecosystem reengineering results in, at least, three specific problems from three different knowledge management (KM) perspectives \cite{2013-KM,skm}. The first issue, from the human-KM interaction perspective, is that without any dedicated conceptual modelling infrastructure, there would only be an \emph{opaque understanding} of information ecosystems vis-à-vis humans \cite{1992-Telos} who interact with the ecosystem in different roles. This results in an \emph{``insufficient understanding (...) of top management in relation to reengineering"} \cite{1998-REENGG}. The second issue, instead, is concerned with the potential problems in clarifying the technological basis of the reengineering process. This primarily stems from, quoting \cite{1998-REENGG}, an \emph{``insufficient understanding about existing data, applications, and IT"} resulting in the subsequent failure to appropriately exploit information technology enablers and capabilities to concretely model and effectuate the reengineering decisions. Thirdly, and perhaps the most challenging problem, concerns delineating the reengineering process itself. Most often than not, the reengineering process is inappropriately defined resulting in it being \emph{``too radical"} for different (levels of) actors involved in effectuating the information ecosystem reengineering. The paper also argues that the aforementioned issues are instantiations of the more fundamental issue of \emph{Conceptual Entanglement} unavoidable in any conceptually-intensive (re)engineering activity.

The paper proposes a \emph{generic} conceptual methodology - the \emph{Representation Disentanglement} approach - as a unified solution to the three aforementioned problems in Information Ecosystem Reengineering. It is a multi-level conceptual modelling mechanism for supporting the formulation, modelling and effectuation of information ecosystem re-engineering decisions founded in the theory and practice of knowledge representation \cite{kr} and \emph{ontology-driven conceptual models} \cite{SCM}. There are three broad justifications in support of an ontology-driven conceptual modelling framework for information ecosystem reengineering. Firstly, the fact that ontology-driven conceptual models provide an \emph{established} means of not only transparently elucidating information ecosystems - entities and their inter-relationships - but also provide an effective \emph{cartographic} solution \cite{2019-Bagchi,2019-ISIIIM} as to the components (e.g., sources, services or systems) that require reengineering. Secondly, the framework can be seen as the starting point for future exploitation of state-of-the-art \emph{semantic modelling methodologies} (see \cite{2021-KO,2022-ER,phd} for such an example) which can guide the development of the technological basis of the reengineering process. Thirdly, due to the two aforementioned justifications, the consequence that such an approach produces machine-processable reengineering models which are \emph{not only} amenable for the reengineering task at hand \emph{but also} can be seamlessly \emph{reused and modified} later on, thus, provisioning an informed bootsrap for future information ecosystem reengineering decisions.

The \emph{novelty} of the proposed approach is \emph{three-fold}. The first highlight is the fact that the knowledge representation and conceptual modelling infrastructure provided by the proposed approach facilitates \emph{explicit linkage} of activities in the reengineering workflow, thus, in effect, exposing the necessary and sufficient reasons for reengineering (as pointed out in \cite{1998-REENGG}). Secondly, and concurrent with the first novelty, the \emph{graph-theoretic} grounding \cite{1983-GT} of the proposed approach-based reengineering workflow renders it \emph{client and management friendly} and completely tailored to their purpose (for instance, purpose being modelled via competency questions \cite{CQ} or cognitive generative questions \cite{2021-KO}). Finally, at a more high-level, an ontology-driven conceptual modelling approach also increases the effectiveness and efficiency of the reengineering workflow being implemented by several folds (the lack of which in state-of-the-art information ecosystem engineering was repeatedly stressed in \cite{1998-REENGG}). 

Notice also, at the very outset, that the focus of the current work is on developing the theoretical basis and advancing a conceptual approach for disentanglement of public sector information ecosystem reengineering decisions grounded in the multi-level theory of conceptual entanglement and disentanglement \cite{2022-ONTOBRAS,IRCDL}. To that end, the paper motivates and justifies the need for each step of the proposed approach with an illustrative exemplification of the decisions concerning reengineering of the information ecosystem of an information services chatbot. It does not, in alignment with the aforementioned focus, concentrate on a detailed practical case study of the approach (which will form the basis of a future research paper). However, the authors are encouraged to consult the cumulative research reported in \cite{2022-ONTOBRAS,IRCDL,ore,ltm} which presents more detailed practical cases in the reengineering of a public sector information ecosystems in geospatial, tourism and healthcare domains following the theory and approach proposed in the current work. Another recent research study \cite{2023-KO}, though focused on a more pedagogical perspective, also includes a detailed elucidation on the implementation of the proposed information ecosystem reengineering framework in the higher education and academic research domain within the public healthcare sector.

The remainder of the paper is organized as follows: In Section (2), the relevant scientific approaches to facilitate information (eco)system reengineering tasks are briefly reviewed. Section (3) details the notion of information ecosystem reengineering introduced with its core characteristics. Section (4) examines ontology-driven conceptual modelling in detail, elucidating their various kinds and established potentiality. Section (5) details a first version of the proposed information ecosystem reengineering approach, the ontology-driven \emph{Representation Disentanglement Approach} for Information Ecosystem Reengineering, with exemplifications on how it can be instantiated and exploited. Notice that the description of the approach is inherently \emph{conceptual} and a detailed technical elucidation is planned as a next paper. Section (6) concludes the paper with summative observations. Given the rapidly expanding role of data and AI in governance, the ability to semantically model and reengineer public sector information ecosystems is no longer optional but \textit{essential}. The proposed approach of Representation Disentanglement, in this context, offers a principled foundation for addressing the design and coordination challenges across multi-stakeholder public sector information governance ecosystems.

\section{Literature Review}
\label{S06}
The literature review is sequentially organized around three broad dimensions. Firstly, general approaches, frameworks and methodologies towards information (eco)system (re)engineering are reviewed. Notice that the spectrum of approaches considered for this dimension are technologically non-intensive in nature. Orthogonal to the first dimension, the second dimension discusses state of the art works in Information and Communication Technology (ICT) and Artificial Intelligence (AI) induced approaches to deal with information (eco)system (re)engineering. Finally, the third section deals with the applications (or the lack thereof) of semantic knowledge management technologies such as ontologies and conceptual models towards information (eco)system (re)engineering.

The early work in \cite{IER1} focused on implementation issues and organization changes concerning the reengineering of an information ecosystem instead of being overwhelmed by the more common study of the effect of information technology on such efforts. The study in \cite{IER2}, instead, described the concrete effort of reengineering information systems at Cincinnati Milacron from a high-level enterprise information planning perspective. The model of information system domains methodology as a novel approach towards information system and ecosystem reengineering was proposed by \cite{IER3}. The work in \cite{IER4} is extremely crucial as it demonstrates that despite the significant involvement and investment of ICT in information ecosystem reengineering, such efforts are failing to take off due to \emph{``the absence of a strong theoretical foundation to assess and analyze"} them. The work in \cite{IER5} offers an interesting foundational take on the aspect of reengineering legacy information sources, services and systems. To streamline requirements engineering in general but also specifically for legacy information source, system and service reengineering, the work in \cite{IER6} proposed a three phased semiotics-driven approach.

Let us now review some of the ICT and AI applications towards information system and ecosystem reengineering. The work in \cite{TIER2} provided an exhaustive overview of a wide range of perspectives and technical architectures in reengineering enterprise information sources, systems and services. The work in \cite{TIER1}, instead, provides a high-level overview of some of the more recent ICT tools, techniques and methodologies currently in vogue in information system and ecosystem reengineering. The case study detailed in \cite{TIER3} and \cite{TIER4} concerns the technological architecture employed and the challenges faced in reengineering information sources, services and systems in logistics and quantum computing systems setting, respectively. From an AI perspective, the study in \cite{TIER5} detailed the reengineering of an information ecosystem dedicated for clinical decision support. The work in \cite{TIER6} described the interesting case of reengineering an expert system founded in the AI technology of machine learning.

Finally, we review applications of knowledge representation and management technologies such as ontologies and conceptual models in information (eco)system (re)engineering. The work in \cite{OIER1} evaluated, from a semantically shallow perspective, the evolution of requirements when shifting from a conventional to a more component modelling based approach in information source, system and service reengineering. An ontology-based approach for the maintenance, repair and overhaul of the reengineering of an information ecosystem in aviation industry was described in \cite{OIER2}. The review in \cite{OIER3} presents an all-encompassing perspective on the application of ontology-driven conceptual modelling techniques in software reengineering, including information ecosystem reengineering. The research reported in \cite{OIER4} points out the high-level problems encountered in an information ecosystem reengineering exercise, e.g., the high degree of complexity, the interchange of action and decision sequences and the dynamic adjustment of information flows in and within such sequences. Finally, in sync with and as a consequence of the above, the work in \cite{OIER5} also pointed out the how an effective management of the aforementioned complexities can affect the \textit{cost, quality, service, and speed} of an information ecosystem reengineering exercise.

Notice that, as per the aforementioned overviews of the different dimensions in research literature, the applications of ontology-driven conceptual models in information ecosystem reengineering are rather \emph{limited}, and, when some form of ontology is employed, there is no underlying general methodological approach to support its implementation and potential reuse. In fact, none of the above research papers factor in and unifies, in a single methodological approach, \textit{all} the following crucial dimensions:
\begin{enumerate}
    \item \textit{Perception}, which plays an early key role in elicitating the key reengineering commitments to be given principle focus in the reengineering exercise, thereby, eliminating perceptual ambiguity;
    \item \textit{Language}, which plays a crucial role in formulating a common and shared vocabulary which can be exploited to effectively intercommunicate information about the shared perception, thereby, eliminating linguistic ambiguity;
    \item \textit{Ontology}, which, even with a shared perception and language, is key to formalize the exact nature of the similarity/differences between the concepts relevant for the reengineering exercise, thereby, eliminating the ontological ambiguity;
    \item \textit{Hierarchy} which, as a consequence of a shared perception, language and  ontology, is key to assert the hierarchical dependence of the concepts relevant for the reengineering exercise, thereby, providing a clear executional view of the information ecosystem reengineering strategy and eliminating hierarchical ambiguity, and,
    \item \textit{Intensionality} which, given all the above layers, is key to model the attribute-level details of reengineering which should be effectuated with respect to each concept in the hierarchy, thereby, eliminating intensional ambiguity.
\end{enumerate}

The current research work is an attempt to plug the research gap as evidenced above, and, to that end, to provide a theoretical basis and advance a conceptual approach to tackle the complexity of the above dimensional layers within a single approach with relevant examples to the original public sector motivating example. Let us now concentrate on the principle dimensions of an information ecosystem reengineering effort and the decisions crucial to each individual dimension.

\section{Reengineering an Information Ecosystem}
\label{S02}
Information Ecosystem Reengineering, as briefly introduced in section (1), is an \emph{umbrella concept} encompassing potential reengineering of (any combination of) information sources, services and systems constituting an information ecosystem in response to \emph{evolving needs} within the \emph{environment} in which the ecosystem functions. There are two concrete observations which need to be highlighted at this stage. Firstly, as aforementioned, the environment which is informationally enmeshed with the information ecosystem is the \emph{causal source} \cite{1998-causality} for re-engineering, i.e. both the \emph{cause} and the resultant \emph{effect} is effectuated vis-à-vis the environment (such as the university knowledge centre from the example in section (1)). Secondly, the evolving needs are crucial given that they provide the requirements \cite{2000-RE} for gauging (quantitatively as well as qualitatively) the features which need to be reengineered. What follows is a brief, \emph{generic} elucidation of the nature of what reengineering information sources, services and systems (individually as well as symbiotically) can be expected to undergo, as an information ecosystem evolves.

Let us first concentrate on the dimension of reengineering of information sources \cite{2013-ISSS} of an information ecosystem, the origin and computational nature of which can be very diverse. A rudimentary source of information can be user inputs via an appropriate application interface or web forms. Web crawling and Application Programming Interfaces (APIs) \cite{API} are also popular means via which information ecosystems source information programmatically. Further, the enterprise-specific or domain-specific databases and knowledge bases (KBs) can also be input sources of information to an information ecosystem. The reengineering of information sources entails either a total shift or partial addition/subtraction of information sources given the change in the imminent purpose which the information ecosystem serves (for instance, from the motivating example, a new need for the information services chatbot to serve high energy physics researchers).

Given the \emph{status quo} of the information sources of an information ecosystem, the design dialogue now shifts to the reengineering of information services. Information services, in general, can again be very diverse in their informational offering. Such services  are conventionally descriptive, analytical or multimedia in nature. However, there are three other kinds of information services which are also emerging to be increasingly important in the context of modern information ecosystems. The first of such kind are the \emph{on-the-go data services} which, based upon a query, harvests and analyzes data on the fly and returns quantified analysis either numerically or via visualizations (with appropriate metadata \cite{SRR125} being a crucial factor in unveiling their meaning). Another kind of such services are the so-called rerouting services which can be seen as a conceptual extension of the traditional referral services which are key to a traditional knowledge resource centre. The third kind of services which are also steadily gaining importance in the face of, for instance, the debate around digital well-being \cite{2021-DWB}, are digital alerts. The strategy for reengineering all these aforementioned information services (both traditional and emerging) can either be due to standalone needs (for example, providing a new form of on-the-go analysis from existing information sources) or emergent needs (for example, a totally novel information service as a resultant of a new genre of information source and/or need).

Thirdly, information system reengineering can be of three major kinds - technological reengineering, component reengineering or architecture reengineering. The first kind, technological reengineering, is relatively lightweight involving the reconditioning of the codebase of the target \emph{`architectural'} component from one technical formalism to another while maintaining its design essence (e.g., reconditioning a KIF KB \cite{KIF} into an OWL KB \cite{OWL2}). Component reengineering usually entails reworking of the design and the functionality of one or more components of the target information ecosystem in response to emerging requirements (e.g., changing the schema of the KB itself to adapt to new requirements). Lastly, architecture reengineering involves a total (or, at least, significant) overhaul of the architecture of the entire information ecosystem potentially, but not only, due to (various combinations of) emergent requirements for newer functionality and technological \emph{passé}. For instance, changing, \emph{en bloc}, the architecture of a chatbot from rule-based paradigm to conversational artificial intelligence based paradigm would fit the scope of architecture reengineering.

Finally, let us examine the nature of the symbiotic interactionism which grounds the aforementioned components of any potential reengineering  of an information ecosystem. Firstly, there is a \emph{many-to-many mapping} between the reengineering of information sources and information services. For instance, the imminent requirement to factor in, say, a couple of new information sources might enforce the reengineering of the information services provided by the chatbot, which, in due time, will also scale the same service beyond the original `couple' of information sources which motivated the reengineering. Secondly, there is also a \emph{many-to-many mapping} between the reengineering of information services and information systems. For example, the need to accommodate a couple of new, earmarked information services might culminate in proposals of varying degrees of reengineering the information system, which, in turn, will open the avenue for more newer genres of information services. Therefore, as a logical consequence to the many-to-many mapping between information sources and services, and information services and systems separately, an inevitable \emph{many-to-many mapping} arises between information sources and information systems which complicates further the overall reengineering exercise.

Notice that the aforementioned practical issues which arise while deciding on an information ecosystem reengineering mechanism are, \emph{non-trivially}, grounded in the more fundamental, layered issue of \emph{representation entanglement} \cite{2022-ONTOBRAS} which is \emph{ubiquitous} in any conceptually-intensive (re)engineering activity. Briefly, representation entanglement arises due to the fact that (human) conceptualizations are causally generated from (human) experientiality, for instance, about a reengineering strategy, and is stratified in nature \cite{2022-ONTOBRAS}.  The first level of entanglement is generated due to the different ways in which different concepts of a target reality (e.g., information reengineering of a chatbot) can be \emph{perceived}. The second level of entanglement arises due to the different ways in which different concepts perceived can be \emph{named}. This is \emph{crucial} given the fact that in any effort of reengineering an information ecosystem, it is necessary to intercommunicate in an \emph{unambiguous vocabulary}. The third level of entanglement pertains to the different \emph{ontological distinctions} \cite{2006-ODT} into which each of such named concepts can be semantically aligned and constrained to. This is crucial in reengineering due to, for instance, the unambiguous conceptual differentiation between an \emph{activity} and the (combination of) \emph{processes} effectuating an activity. The fourth level of entanglement, given the ontological distinction alignment, instantiate as the different ways in which the concepts can be \emph{hierarchically interrelated}, this being perhaps the \emph{most important} step in \emph{cartographically visualizing} the reengineering activity. The fifth and the final level of entanglement occurs due to the different ways in which different concepts in the hierarchical model can be \emph{intensionally characterized}, i.e., associated to relevant \emph{attributes}. This is \emph{key} to an unambiguous, parametric explanation behind reengineering a specific component (represented as a \emph{concept} in the hierarchical conceptual model). Before moving on to the proposed \emph{Representation Disentanglement} approach which is a major advance towards \emph{disentangling} the aforementioned multivariate mappings which \emph{entangle} reengineering processes, let us briefly examine the foundations and technical advantages offered by ontology-driven conceptual models.

\section{Ontology-Driven Conceptual Modelling}
\label{S03}
The employment of models and modelling paradigms for abstract representation of (any fragment of) reality has been an established and pervasive practice in modern science and engineering \cite{2022-model}. Conceptual models are models which elucidate \emph{``some aspects of the physical and social world around us for purposes of understanding and communication"} \cite{1992-Telos} and conceptual modelling refers to the activity which focuses on theoretically and implementationally well-founded conceptual model design. Let us elaborate the notion of `conceptual models' by examining four of its key tenets in ascending order of concretization (as alluded to in \cite{2020-GGMCM}). 

Firstly, conceptual models are \emph{mental models} comprised of mental representational primitives (\emph{aka} mental concepts) via which one filters and internalizes a \emph{relevant} fragment of reality (e.g., a specific case of reengineering). Secondly, as a direct consequence of the fact that conceptual models are cognitively grounded \cite{2022-ONTOBRAS}, they are also intrinsically \emph{intensional} which, in simple terms, mean they \emph{refer to} or \emph{are about} something. Thirdly, conceptual models have mandatory \emph{conceptual semantics} which means, in other words, the concepts they (mentally) encode aren't merely \emph{logical} in nature but \emph{definitively} have a concrete or abstract referrent. The second and the third tenets are crucial for information ecosystem reengineering because of the \emph{symbol grounding problem} \cite{1990-SGP} which guarantees the uselessness of any model which is \emph{intensionally} and \emph{semantically} disconnected from the target reality it models. Finally, the fact that any conceptual model is a \emph{computational independent model} necessitating the observation that it can be \emph{formally} instantiated via any recognized conceptual modelling languages. One famous example of such a modelling language is \emph{Telos} \cite{1990-telos,1992-Telos} which, though focused on knowledge representation-intensive conceptual modelling for information systems, holds in general for any application arena.

Given the backdrop of conceptual modelling in general, let us now concentrate on one of its most established kinds - ontology-driven conceptual modelling \cite{2002-ODCM}, which is grounded in the theory of formal ontology \cite{1995-ontologies}. Formal ontology has been a \emph{tried-and-tested} tool to encode and exploit knowledge (at different levels of abstraction) in the arena of artificial intelligence. It has been defined as \emph{``a formal, explicit speciﬁcation of a shared conceptualization"} \cite{1998-Studer}, thus, ensuring that \emph{all} of its instantiations satisfy the aforementioned foundational tenets of conceptual models. Additionally, ontology-driven conceptual models are marked by \emph{at least} four extra dimensions. Firstly, the concepts which a domain ontology represent should \emph{ideally} conform to \emph{top-level} philosophical distinctions such as \emph{identity} (how to uniquely recognize a concept) and \emph{unity} (how such a concept is constituted) \cite{2002-OC}. The aforementioned conformances result in ontology-driven conceptual models being highly intra-domain \emph{interoperable} and consequently embody enhanced potential for being \emph{reused}. For example, an ontology-driven conceptual model encoding the reengineering of an information services chatbot \emph{X} can later be \emph{reused and modified} as a template for the reengineeering of an information services chatbot \emph{Y}. Secondly, ontology-driven conceptual models interlink the concepts using \emph{object properties} \cite{OD101} to form a directed acyclic graph \cite{1983-GT}. Thirdly, each individual concept in the ontology-driven conceptual model can be further described with what are referred to as \emph{data properties} \cite{OD101} (also referred to as \emph{attributes}). Finally, for ontology-driven conceptual models, the entire model can be annotated with \emph{metadata} via \emph{annotation properties}. This is pivotal from several perspectives, e.g., recording the \emph{provenance} of information ecosystem reengineering, etc.

Let us now concentrate on the established ontology-driven conceptual modelling methodologies and their associated technological standards. The survey in \cite{OEM-IJCAI} provides a synopsis of early generation ontology-driven conceptual modelling methodologies which were designed by adapting parameters from software life cycle process engineering. Methontology \cite{METHONTOLOGY} proposed a \emph{``life cycle to build ontologies based in evolving prototypes"} and founded its strategy towards developing domain-specific ontology-driven conceptual models. Ontology Development 101 \cite{OD101}, instead, offered the flexibility of choosing top-down, bottom-up or middle-out approaches in engineering ontology-driven conceptual models. More recently, the NeOn methodology \cite{NeOn} offers a set of \emph{very generic} scenarios for reuse, re-engineering and merging of ontological resources. Orthogonal to the methodologies above, the works in \cite{FOCAKR,UBSO} adopt a philosophically-grounded approach in designing ontology-driven conceptual models. In addition to the aforementioned methodologies, ontology-driven conceptual models are also backed by an increasing cohort of standard technologies (e.g., the entire stack of semantic web standards by the W3C consortium \cite{2010-W3C}).  

We now elucidate the step-wise \emph{Representation Disentanglement} approach to information ecosystem reengineering which builds upon the theory and methodological practice of ontology-driven conceptual models as described above.

\section{The \emph{Representation Disentanglement} Approach}
\label{S04}
Given the background about the issues concerning the reengineering of an information ecosystem, the paper proposes \emph{Representation Disentanglement} (see Table \ref{RDL} for a summative overview) as a generic solution approach founded in ontology-driven conceptual modelling \cite{2002-ODCM,SCM,GG}. It is essentially a multi-level conceptual modelling strategy to tackle representation entanglement that instantiates in information ecosystem reengineering (see section \ref{S02}). To that effect, the representation disentanglement approach proposes a set of very generic guiding \emph{normative principles} which, if considered as best practice for each of the five levels, can enforce disentanglement of the representation entanglement while at the same time, provide an \emph{unambiguous} and \emph{standardized} \emph{knowledge management} basis on which different actors engaged in the information ecosystem reengineering decision chain can communicate. The concrete strategy for disentanglement for each individual level is discussed as follows.

\begin{table}[ht]
\centering
\caption{Summary of the Representation Disentanglement layers in Information Ecosystem Reengineering}
\begin{tabular}{|p{3cm}|p{3cm}|p{3cm}|p{3cm}|}
\hline
\textbf{Representation Level} & \textbf{Key Objective} & \textbf{Activity} & \textbf{Fix Ambiguity} \\
\hline
\textbf{1. Perception} & Identify what is perceived and what matters & Define target reality; elicit viewpoints & Perceptual ambiguity \\
\hline
\textbf{2. Labeling} & Standardize how concepts are named & Fix controlled vocabulary; assign labels and IDs & Linguistic ambiguity \\
\hline
\textbf{3. Ontology} & Classify what each concept \emph{is} & Apply ontological meta-properties (e.g., rigidity, identity) & Ontological ambiguity \\
\hline
\textbf{4. Hierarchy} & Organize concepts into classificatory structure & Build a taxonomy using classification principles & Hierarchical ambiguity \\
\hline
\textbf{5. Intensionality} & Define attributes and relationships & Specify data and object properties & Intensional (attribute-level) ambiguity \\
\hline
\end{tabular}
\label{RDL}
\end{table}

\textbf{\emph{Perception (Disentanglement):}} Let us first concentrate on the norms which tackle the entanglement instantiated due to the very nature of the \emph{Perception}. The approach recommends the sequential fixation of the following:
\begin{itemize}
    \item Firstly, the \emph{target reality} (e.g., reengineering of the information services chatbot) should be precisely delineated with respect to their \emph{spatio-temporal} extent. In the case of a distribution of several smaller component realities (e.g., phase-by-phase reengineering of the chatbot), the target reality should be modelled as a disjoint union of the component realities (i.e., of component spatio-temporal extents). Notice that such a spatio-temporal delineation can be as \emph{general or specific} as possible depending upon the information ecosystem requirements elucidated, for instance, via Competency Questions (CQs) \cite{CQ}.
    \item Given the delineation of the target reality, the second activity should determine the concepts (e.g., the different components and their sub-components, activities, etc.) which requires to be modelled within the chosen target reality and the \emph{viewpoints} to be considered while modelling them. Noticeably, the technique of intra-organizational \emph{focus groups} \cite{1996-FG} can be employed to harmonize and accommodate diverse viewpoints concerning the information ecosystem reengineering decision-making. 
\end{itemize}

\noindent The two aforementioned norms facilitates selection of the \emph{intended ontological commitment} (i.e., the body of concepts to be considered) of the target reality very precisely and thus avoids instances of overcommitment and undercommitment which frequently arise in every domain (see \cite{SCM} for related exemplifications). This level facilitates disambiguation of the \emph{perceptual ambiguity} that is imminent in the initial phase of any information ecosystem reengineering task and thereby, helps disentangle the very perception of the task.

\textbf{\emph{Labelling (Disentanglement):}} Given the disentanglement at the perception level, the reengineering consideration shifts focus on the guiding norms for disentangling the entanglement occurrent in the \emph{Labelling} level due to language related issues. The norms are as follows:
\begin{itemize}
    \item Firstly, the fixation of the underlying natural languages(s) and the \emph{controlled vocabulary}, the terminology of which has wide \emph{inter-labeller agreement} and can be used to uniquely name the perceived concepts. Notice that the above fixation is extremely crucial in order to achieve a \emph{common and interoperable vocabulary} in order to facilitate disambiguous inter-communication amongst the different actors involved in the information ecosystem reengineering task. International terminological standards for various domains (e.g., ISO terminology for information management and reengineering) can be exploited for this purpose. Such a choice forces \emph{labelling disambiguation} out of the multiplicity of possible labellings in the selected language(s) on one hand, and absolves the effect of linguistically-grounded labelling conflicts on the other hand.
    \item Optionally, given the fixation of the common terminological base in the previous step, the next step, especially key in scenarios like cross-departmental communication and geographically distributed information transfer about the reengineering activity, is to further \emph{disambiguate} the uniquely labelled concepts via associating to each of such concepts a unique alphanumeric identifer (such as the codes provided by an appropriate general purpose and/or a domain-specific information dictionary).
\end{itemize}

\noindent Overall, this level facilitates disambiguation of the \emph{linguistic ambiguity} that is imminent after the perceptual consolidation of any information ecosystem reengineering task and thereby, helps disentangle and formulate a common and shared language which can be exploited to effectively communicate information about reengineering task-related decision.

\textbf{\emph{Ontological Alignment (Disentanglement):}} Once the concepts are encoded via a disambiguated label, the next key step is to perform an \emph{ontological analysis} with respect to each of the labelled concepts from the previous level (for example, the components of a chatbot to be reengineered and the corresponding activities and processes to be implemented to do so), this being the guiding norm to disambiguate the ontological nature of each such concept. Ontological analysis usually employs a set of metaproperties to \emph{``characterize relevant aspects of the intended meaning of the properties, classes, and relations that make up an ontology"}\cite{2002-OC}. To take an example, in the OntoClean framework \cite{2002-OC} which is employed for ontological analysis in diverse activity arenas including reengineering, the notion of \emph{essence} and its special case of \emph{rigidity} is crucial in ascertaining the \emph{ontological stance} of, for example, the reengineering concepts (e.g., components, processes, activities, tasks, etc.). Similarly, analysis of concepts from the perspective of ontological \emph{identity}, \emph{unity}, \emph{endurance, perdurance} etc. facilitates, at a later stage, development of ontologically well-founded conceptual models. Thus, the cumulative target of the ontological analysis performed at this level is disambiguation of the \emph{ontological ambiguity} inherent even in a common, shared language, i.e., for example, distinguishing \emph{what is} a component, an activity and/or a process in an information ecosystem reengineering task, and \emph{how} they should be existentially and functionally differentiated.

\textbf{\emph{Hierarchical Modelling (Disentanglement):}}
Given the ontological disambiguation of the labelled reengineering concepts, the next step is to create the taxonomical graph of concepts which encodes how the various concepts (e.g., components, subcomponents, activities, processes, tasks, etc.) hierarchically interlink with each other, thereby, delineating the order in which a reengineering is to be componentially implemented. To build such a disentangled taxonomical hierarchy out of ontologically analysed concepts, the proposed approach \emph{adapts} the hierarchy modelling steps enshrined in the \emph{four-step} ontologically well-founded classification theory by Ranganathan \cite{srr67}. The first step concerns the selection of a differentiating \emph{characteristic} for classification of a concepts at a single level in the hierarchy. The second step involves the \emph{succession of characteristics} which, in other words, involve the selection of characteristics for classification of concepts at each successive level in the hierarchy. The third step involves how \emph{sibling concepts} should be modelled within a single level in the hierarchy (termed \emph{array} in \cite{srr67}). The fourth and the final step concentrates on the ontological consistency of (each of) the \emph{single path} in the ontological hierarchy (termed \emph{chain} in \cite{srr67}). The guiding normative principles (termed \emph{canons}) for each of the above four steps are briefly mentioned as follows:

\begin{itemize}
\item In the first step, the entanglement in the selection of the differentiating characteristic is disentangled by exploiting the canons of \emph{relevance} (stating that such a characteristic should be relevant to the purpose at hand) and \emph{ascertainability} (stating that such a characteristic should be perceptually ascertainable). For example, in the case of building an ontology-driven conceptual model for reengineering the information services chatbot, the first classification characteristic can be fixed to be that of the \emph{constituent of reengineering}, for example, generating sub-concepts such as information \emph{ources, services} and \emph{systems}.
\item The entanglement in the second step of choosing the succession of characteristics is disentangled by employing the canon of \emph{relevant succession} which enforces that the selection of successive differentiating characteristics across the depths of a taxonomy should be founded solely on purpose. For instance, the second characteristic for the above ontology-driven conceptual model can be, for instance, for the reengineering of \emph{information services}, the \emph{type of information services} generating, for example, sub-concepts such as \emph{synchronous} and \emph{asynchronous} information services.
\item The canon of \emph{exhaustiveness} is employed to disentangle the entanglement in arrays by ensuring that all the concepts at a specific depth in the taxonomic hierarchy are exhaustively classified at the next depth and thereby, additionally ensuring the exclusivity of chosen purpose-driven differentiating characteristic(s).
\item Finally, the entanglement for modelling a chain is eliminated via the canon of \emph{modulation} which ensures that there are no missing \emph{conceptual} links in any possible path of a taxonomy. For example, this canon ensures that all the paths in the hierarchy of information services to be reengineered are populated by concepts at all depths and rules out missing links (which otherwise hints at a many-to-many crossover in the succession of characteristics).
\end{itemize}

\noindent This step is perhaps the most crucial sub-step which compose the overall strategy of \emph{Representation Disentanglement} due to two specific reasons. Firstly, it facilitates the disentanglement of the many-to-many possibilities of conceptualizing the information ecosystem reengineering task (each of which means different conceptual hierarchies). Thus, it helps in deciding a \emph{common and shared} conceptualization of the reengineering decisions. Moreover, secondly, it facilitates disambiguation of such \emph{hierarchical ambiguity} via a highly visually cartographic means. For example, the entire reengineering conceptual hierarchy can be easily visualized using any of the different flavours of visual grammars that are employed in mainstream conceptual modelling. This level of conceptual clarity and hierarchical modeling is especially vital in public sector governance applications, wherein, decisions about information ecosystem transformations must be transparent, traceable and explainable to various oversight bodies, stakeholders and the public at large. The ability to make explicit the rationale, scope and dependencies of reengineering decisions support algorithmic accountability and institutional auditability which are key principles in emerging public sector AI ethics guidelines.

\textbf{\emph{Intensional Definition (Disentanglement):}} Given the norm-based disentanglement of the hierarchical taxonomic model of the reengineering concepts, the final entanglement at the intensional definition level is fixed by precisely determining the relations (object properties) and attributes (data properties) out of the manifold property interlinkages and combinations that can be conceptualized for a particular information ecosystem reengineering task. This facilitates disambiguation of the \emph{intensional ambiguity} explicit in the hierarchical graph that was designed in the previous step. Notice that both the \emph{conceptual interlinkage} and the \emph{conceptual description} is key for any information ecosystem reengineering effort. The former delineates and simplifies the complexity of the conceptual connections which drive the reengineering effort while the latter faciliates encoding the metadata about the pre-engineered and post-engineered state of any component that is reengineerd. Notice that this final ontology-driven conceptual graph can be computationally modelled and visualized as a \emph{machine readable} ontology, for instance, in the Web Ontology Language (\texttt{OWL})\footnote{https://www.w3.org/TR/owl2-primer/}, thereby, keeping open the future possibility of its \emph{reuse and extension} as an information ecosystem reengineering template.


\section{Conclusion}
\label{S07}
The paper elucidated in detail the representation entanglement \emph{implicit} and \emph{ubiquitous} in information ecosystem reengineering decision-making and proposed a multi-level ontology-driven conceptual modelling strategy grounded in \emph{Representation Disentanglement} to absolve such entanglement and make \emph{explicit} the decisions adopted. In future, the efficacy of the conceptual framework developed and elucidated in this paper will be tested and possibly enhanced by experimenting with more concrete use cases in different domains. Another line of research can be the detailed study of the proposed \emph{Representation Disentanglement} approach in information ecosystem reengineering from an ablationary perspective increasingly relevant in smart knowledge management \cite{2019-SKM} scenarios. Further, some of the future applications of the proposed approach may extend to digital public sector information and governance ecosystems, including smart city knowledge infrastructures, participatory open government platforms and adaptive policy support systems. In such contexts, where information ecosystems mediate critical services and multi-actor interactions, the proposed approach of representation disentanglement can offer a robust strategy for sustaining semantic clarity, accountability, and coherence in continuous reengineering efforts.

\bibliographystyle{unsrt}
\bibliography{IJKMS}

\end{document}